# Framework of Resilient Transmission Network Reconfiguration Considering Cyber-Attacks

Chao Yang, *Member, IEEE*, Gaoqi Liang, *Member, IEEE*, Steven R. Weller, *Member, IEEE*, Shaoyan Li, *Member, IEEE*, Junhua Zhao, *Senior Member, IEEE*, and Zhaoyang Dong, *Fellow, IEEE*

*Abstract*—Fast and reliable transmission network reconfiguration is critical in improving power grid resilience to cyber-attacks. If the network reconfiguration following cyber-attacks is imperfect, secondary incidents may delay or interrupt post-attack restoration of the power grid. This paper proposes a framework of resilient transmission network reconfiguration, taking into account the impacts of cyber-attacks in the network reconfiguration process. First, the mechanism of cyber-attack propagation is analyzed based on the characteristics of network reconfiguration. Second, systematic resilience indices are specially extracted in which the impact of cyber-attacks on network reconfiguration is quantified. These indices are defined in terms of the restoration characteristics of the transmission power system. Third, representative cyber-attack incidents motivate an optimization-based model of resilient transmission network reconfiguration, and an optimal reconstruction scheme is obtained. Finally, simulation results based on the IEEE 39-bus system verify the feasibility and effectiveness of the proposed framework in enhancing power grid resilience to cyber-attacks.

## I. Introduction

WITH the development of smart grids, the modern power system has become a cyber-physical power system (CPPS). While sensing, computation and actuation in the grid dramatically enhance self-healing capabilities, vulnerability to cyber-attack introduces new possibilities for malicious activity.

In recent years, blackouts of power systems caused by cyber-attack have occurred on a number of occasions. The most well-known attack is the 2015 Ukraine blackout [1] caused by synchronized and coordinated cyber-attacks, resulting in approximately 225,000 customers suffering from power outages for hours. The most recent blackout caused by cyber-attacks occurred in Venezuela [2] in 2019. This attack interrupted the power supply in eighteen states for several days, leading to enormous losses to the national economy and a great impact on people's lives, livelihoods and social stability.

In the CPPS, attackers steal the confidential operation data without permission and use security flaws to attack the control system or its resources. Common forms of cyber-attack in the literature include the denial-of-service attack, false data injection attack [3], man-in-the-middle attack, replay attack, cyber-topology attack, etc. The impacts of cyber-attacks are primarily reflected in the security and economics of the power system [4]. The resilient power system [5], on the other hand, can effectively mitigate the impacts of cyber-attack incidents. Power system resilience refers to the capacity to forecast, absorb and recover from external, high-impact, low-probability disturbances. Its main properties include resourcefulness, robustness, redundancy, and restorability [6]. The power system restoration [7] is therefore an important research topic. The restoration process of bulk power systems includes three main stages [8]: the black start stage, the transmission network reconfiguration stage and the load restoration stage.

Transmission network reconfiguration [9] is the bridge between the black start stage and the load restoration stage. Its main task is to establish a power-supplied path for generators without self-starting capability and critical loads as soon as possible. Existing methods for path establishment are mostly based on graph theory, such as the shortest path algorithms [10], [11], which makes the final reconfigured network a tree graph. Moreover, the traditional strategy has an implicit assumption that the cyber system is intact and available during the restoration process. The traditional reconfiguration scheme is therefore idealistic.

In reality, however, the cyber system may shut down due to loss of power supply, fault or damage to communication equipment after a blackout, and even cyber-attacks, which will further weaken the restoration capability of the power system [12]. During the restoration process, the CPPS is vulnerable to secondary incidents that may delay the restoration and even lead to secondary blackouts. Current research into network reconfiguration rarely considers secondary incidents, especially malicious cyber-attacks. Thus, this paper proposes a framework of resilient network reconfiguration considering cyber-attacks. The main contributions are as follows:
1) This paper combines cyber-attacks with transmission

This work is partially supported by National Natural Science Foundation of China (72331009, 72171206, 52107092), Shenzhen Key Lab of Crowd Intelligence Empowered Low-Carbon Energy Network (No. ZDSYS20220606100601002), and the Shenzhen Institute of Artificial Intelligence and Robotics for Society (AIRS). (Corresponding author: Junhua Zhao and Gaoqi Liang.)
C. Yang, and J. Zhao are with the School of Science and Engineering, The Chinese University of Hong Kong, Shenzhen, Shenzhen 518100, China (e-mail: yangchao@cuhk.edu.cn; zhaojunhua@cuhk.edu.cn).
G. Liang is with the School of Mechanical Engineering and Automation, Harbin Institute of Technology, Shenzhen, Shenzhen 518055, China (e-mail: gaoqi.liang@outlook.com)
S. R. Weller is with the School of Engineering, University of Newcastle, Callaghan, NSW 2308, Australia (e-mail: steven.weller@newcastle.edu.au).
S. Li is with the School of Electrical and Electronic Engineering, North China Electric Power University, Baoding 071003, China (e-mail: shaoyan.li@ncepu.edu.cn).
Z. Dong is with the School of Electrical and Electronic Engineering, Nanyang Technological University, 639798, Singapore (e-mail: zydong@ieee.org).

network reconfiguration, extending the traditional network reconfiguration by considering the impacts of secondary incidents caused by cyber-attacks.
2) The cyber-attack propagation mechanism in the CPPS, propagation process and impact assessment in the network reconfiguration stage are systematically analyzed.
3) From the perspective of restorability and robustness, a resilience index system is extracted for assessing impacts of cyber-attacks on network reconfiguration dynamically.

The remainder of the paper is organized as follows: Section II proposes a framework of resilient network reconfiguration with cyber-attacks. Section III presents the optimization model of network reconfiguration. Case studies are analyzed in Section IV and conclusions are presented in Section V.

## II. Framework of Resilient Transmission Network Reconfiguration Considering Cyber-attacks

### A. Cyber-attack Propagation Mechanism in the CPPS

#### 1) Cyber-attacks in the CPPS

In CPPS, the cyber system, including all communication equipment and information, is assumed to be the direct target of cyber attackers. Illegal access to information, interference with information transmission (e.g., loss, interruption and delay) and tampering with information (e.g., overwrite, delete, insert, replace) are three common attack vectors. Failure of the cyber system will thereafter affect the security and stability operation of the physical system, most critically the power dispatching automation network module and associated energy management system, distribution management system, wide area monitoring system and advanced metering infrastructure [13]. A successful cyber-attack on one or more of these modules can change the operational state of the power system between the normal state, alert state, emergency state and restoration state, with the impact scope ranging from a single piece of equipment to the entire power system.

#### 2) Cyber-Attack Propagation Mechanism in the CPPS

The cyber-side and physical-side of CPPS have complex and diverse interdependence. The impacts of cyber-attack on power systems can be divided into direct and indirect impacts. Direct impact refers to the failure of one side leading to the misoperation or direct shutdown of equipment on the other side. Indirect impact refers to the failure of one not immediately affecting the other side, but rather weakening the capability of its equipment or systems to resist other disturbances [4].

In this study, a graph theory-based approach is introduced to establish the impact propagation mechanism of cyber-attacks in the CPPS for estimating both direct and indirect impacts of cyber-attacks. As shown in Fig. 1, the cyber system and physical system are simplified into two graphs, wherein nodes in the cyber system represent communication equipment or control centers, branches represent information transmission paths such as optical fibers and power line carriers. Nodes in the physical system represent power plants or substations, branches represent transmission lines. Therefore, the impact propagation process can be given as follows: 1) cyber-attackers attack the nodes and/or branches of the cyber system; 2) based on the spatial mapping relationship, the impacts of cyber incidents are transmitted to the physical system; and 3) failures and incidents of the CPPS thereby result.

As shown in Fig. 1, four main impact propagation processes are included: attack matrix, cyber incidents, initial impacts and secondary impacts.

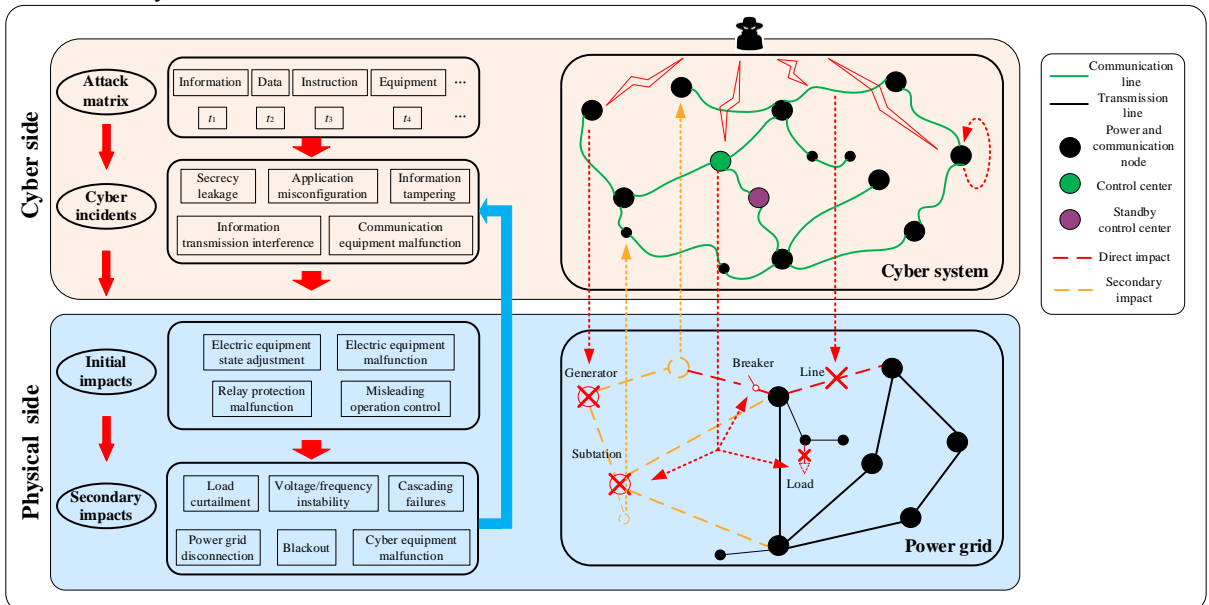

Fig. 1. Cyber-attack propagation mechanism in the CPPS

The attack matrix [14] is the path that attackers can take to gain access to the cyber system, including the attack target (e.g., the information, data, instruction, and equipment) and attack time. Cyber incidents are the ultimate cyber effects caused by cyber-attacks including secrecy leakage, application misconfiguration, information tampering and transmission interference or communication equipment malfunction, etc. It should be clear that a single cyber-attack could lead to many cyber incidents. Likewise, several cyber-attacks may lead to the same cyber incident. Additionally, multiple cyber incidents may happen concurrently or sequentially.

The initial impacts on the physical system are the result of cyber incidents, such as state adjustment or malfunction of electrical equipment, relay protection malfunction or misleading operational control of the power grid. A single cyber incident can lead to many initial impacts, and several cyber incidents may lead to the same initial impact. In addition, some cyber incidents may not immediately have a direct impact on the physical system, such as the fault resulting in a circuit breaker failing to operate, which only results in impacts when it needs to be disconnected. More importantly, the initial impacts may cause secondary impacts subsequently.

The secondary impact is complex and various. It includes the subsequent impact on both the cyber and physical systems. For one thing, the secondary impact on the physical system is caused by secondary failures, which are a series of incidents that occur after initial impacts due to the complex operation correlation of power grids or improper regulation measures of dispatching centers. Generally, the consequences of secondary failures are more serious, which include load curtailment, voltage and frequency stability problems, cascading failures, power grid splitting and blackout. For another, the secondary impact on the physical system can also loop back into new cyber incidents. Disruptions or degradations of power supply may lead to malfunction of cyber equipment, resulting in the functional failure of partial cyber system. However, it should be noted that the backup power supply in communication nodes is capable of providing several hours of energy. A long-term power outage might be required to cause such impacts.

### B. Cyber-Attack Propagation Process in Network Reconfiguration Stage

#### 1) Characteristics of Network Reconfiguration

The typical restoration process is represented by the black curve in Fig. 2. The ideal restoration state is between time $t_r$ and $t_{pr}$, and the restoration level is from $R_b$ to $R_{pr}$. Normally, $R_{pr}$ is lower than the initial level before blackout.

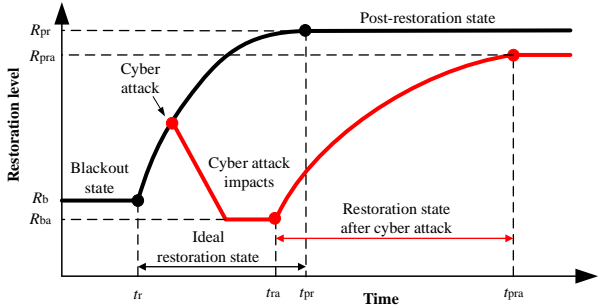

Fig. 2. The typical restoration process with and without cyber-attacks

The main task of network reconfiguration is to establish power-supplied paths for offline generators and important loads as soon as possible, so as to establish a stable skeleton network, and to improve the transmission capacity and system strength. In this stage, the non-black-start unit (NBSU) with the best performance will be restarted in advance. The restored skeleton network is mostly a tree graph with the characteristic of rapidity, but weak strength, resulting in the difficulty of resisting large disturbances or incidents. The restored load mainly plays the role of regulating voltage and balancing generator outputs. Meanwhile, critical loads that are important to social stability and system restoration should be restored quickly.

As the system strength of the restored network is relatively weak, the possibility of secondary incidents increases. If a cyber-attack happens, it will lead to an increase in the number of failed equipment and the power grid may collapse again, so the impact can be more serious. Generally, as shown by the red curve in Fig. 2, the lowest level after re-blackout is lower, $R_{ba}<R_b$, and the restoration time is longer, $(t_{pra}-t_{ra})>(t_{pr}-t_r)$. Moreover, if the power grid collapses again, the restoration level of the post-restoration state is lower, $R_{pra}<R_{pr}$. Therefore, in the network reconfiguration stage, it is not enough to pay attention to rapidity alone, but also resilience improvement.

#### 2) Cyber-attack Propagation Process in Network Reconfiguration Stage

Cyber attackers targeting the network reconfiguration should fully consider the above characteristics, *i.e.*, focusing on critical generators, restoration paths and loads for the purpose of delaying or destroying this stage. First, attackers should only target restored electric equipment, since attacks on shutdown equipment are ineffective. Second, the purpose of attacks is to cause physical impacts and delay/destroy the restoration process. Third, it is better for attackers to maximize the impact of the existing attack capability on the currently restored power grid. In addition, it is difficult for attackers to maximize the attack effect in the overall reconfiguration process, because this process is complex and changeable, attackers hard to fully know all the reconfiguration information, and cyber-attacks cannot affect the artificially controlled equipment.

To analyze the propagation process of cyber-attacks, the reconfiguration process from the black start unit (BSU) to the full network is regarded as the superposition of the sequential steady-state time sections and the dynamic transition process between those sections. As shown in Fig. 3, the network reconfiguration is such a process of switching between different restoration states and spiraling towards the overall reconfiguration. In the initial state, only the BSU is operating. Then, restoration actions are taken for network reconfiguration. Generally, there are several feasible schemes for each restoration step. When the restored power grid experiences cyber-attacks, it will transit to a new stable state with a lower restoration level. Finally, the network reconfiguration can finish even after suffering malicious cyber-attacks.

The common propagation process of cascading failures [15] is introduced to simulate the transit process for the restored grid. First, based on the initial outage caused by cyber-attacks, the 1st new steady-state of the post-attack grid is recalculated. Then, overloaded equipment fails and the 2nd new steady-state is recalculated. By analogy, until there is no overloaded equipment. Finally, all failed equipment is found and the final stable state of the post-attack restored grid is obtained.

### C. Impact Assessment of Cyber-Attacks in Network Reconfiguration Stage

The assessment of cyber-attack impact is analyzed from two perspectives: the overall network reconfiguration process, and the restoration time-section. A resilience index system is proposed for quantitative evaluation as follows.

#### 1) Restorability of the Overall Reconfiguration Process

Restorability is the main resilience feature of the overall reconfiguration process. It refers to the capability of restoring load rapidly by effectively control after a blackout. The main impact of cyber-attacks is to delay the restoration process. Based on the network reconfiguration theory, restorability indices are summarized as follows.

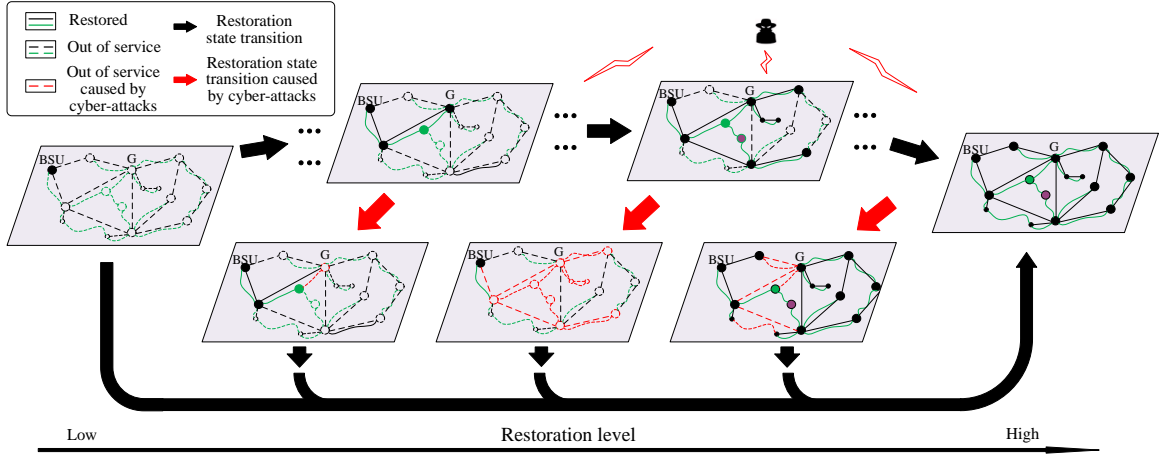

Fig. 3. Process diagram of network reconfiguration considering cyber-attacks.

*(1) The restoration time index*

$$R_1^{res} = T \tag{1}$$

where $T$ is the total restoration time of the overall process.

*(2) The average generator ramping rate index*

$$R_2^{res} = \int_0^T \sum_{g \in G} r_{G,g}(t) dt / T \tag{2}$$

where $r_{G,g}(t)$ is the ramping rate of generator $g$ at time $t$, and $\boldsymbol{G}$ denotes the set of generators.

*(3) The total capacity of restored load index*

$$R_3^{res}(t) = \sum_{l \in L} P_{L,l}(t) \tag{3}$$

where $P_{L,l}(t)$ is the restored load value of load node $l$ at time $t$, and $\boldsymbol{L}$ denotes the set of load nodes.

*(4) The network connectedness for continuous power supply*

$$R_4^{res}(t) = \sum_{s \in G_{on}} \sum_{o \in L_{on}} c_{con,s,o}(t) \bigg/ \sum_{g \in G} \sum_{l \in L} c_{con,g,l} \tag{4}$$

This index means the continuous power supply capability of the restarted generators to restored loads. Where, $c_{con,s,o}(t)$ is the connectedness of the restored grid (between restarted generator $s$ and restored load node $o$), $c_{con,g,l}$ is the connectedness of the initial power grid before blackout (between generator $g$ and load node $h$), $\boldsymbol{G}_{on}$ is the set of restarted generators, and $\boldsymbol{L}_{on}$ is the set of restored load nodes. With the same node number, the larger this index, the higher the continuous power supply capability of reconfigured networks.

*(5) The network redundancy for continuous power supply*

$$R_5^{res}(t) = \frac{1}{N(N-1)} \sum_{i,j \in V(t)} \kappa_{i,j} (i \neq j) \tag{5}$$

This index is expressed by the connectivity of restored power grids, which is defined as the average connectivity of all node pairs in restored grids [16]. Where, $\kappa_{i,j}$ is the minimum number of lines to be removed to make nodes $i$ and $j$ disconnected in restored power grids, $V(t)$ is the set of restored nodes, and $N$ is the total node count in the initial power grid. With the same number of nodes, the larger this index, the higher the redundancy of the restoration path and the better the invulnerability of reconfigured networks.

*2) Robustness of the Restoration Time-section*

The restoration time section refers to the restored partial power grid that operates stably at a restoration time. Its main resilience feature is robustness which refers to the capability to resist accidents and maintain power supply when the restored power grid is disturbed. The impacts of cyber-attacks include multi-fault of equipment, cascading failure or blackout again. Based on the analysis method of *N-k* contingency [17] in the risk assessment theory, the quantitative indices of robustness that reflect the current impacts and the subsequent impact on the restoration process of cyber-attack are proposed as follows.

*(1) The restoration time delay index*

$$R_1^{rob} = \Delta T \tag{6}$$

where $\Delta T$ is the restoration delay time of a network reconfiguration scheme with cyber-attacks.

*(2) The loss of generating capability index*

$$R_2^{rob}(t) = \sum_{g \in G} \Delta r_{G,g}(t) \tag{7}$$

where $\Delta r_{G,g}(t)$ is the difference of ramping rate for generator $g$ after cyber attacks at time $t$.

*(3) The load curtailment index*

$$R_3^{rob}(t) = \sum_{l \in L} \Delta P_{L,l}(t) / R_3^{res}(t) \tag{8}$$

where $\Delta P_{L,l}(t)$ is the load curtailment of load $l$ after cyber-attacks at time $t$.

*(4) The decrease of network connectedness index*

$$R_4^{rob}(t) = \Delta R_4^{res}(t) / R_4^{res}(t) \tag{9}$$

where $\Delta R_4^{rec}$ is the network connectivity variation of the restored power grid after being attacked at time $t$.

*(5) The decrease of network redundancy index*

$$R_5^{rob}(t) = \Delta R_5^{res}(t) / R_5^{res}(t) \tag{10}$$

where $\Delta R_5^{rec}$ is the network redundancy variation of the restored power grid after being attacked at time $t$.

## III. Modeling and Solution of Resilient Transmission Network Reconfiguration Considering Cyber-attacks

### A. Strategy of Resilient Network Reconfiguration

Generally, the main task of the network reconfiguration stage is to restore NBSUs and important loads rapidly. While rapidly restoring, the resilience of the restored power grid should also be enhanced. The resilience enhancement can greatly improve the capability of power grids to deal with disturbances and faults, guarantee the efficient transmission of power energy and lay the foundation for load restoration. Resilience enhancement can be achieved by the following restoration control measures.

1) As for generators, the NBSUs with good restoration and regulation performance should be restarted quickly.
2) As for the network structure, the characteristic of high connectedness and redundancy is essential to reduce the loss caused by the re-fault of restored equipment.
3) As for loads, they should be restored in coordination with the network, so that the power flow distribution is more reasonable to reduce the loss caused by cyber-attacks.

Rapid restoration generally leads to inadequate robustness, whereas sufficient robustness typically leads to slow restoration. When formulating a resilient network reconfiguration strategy, it is therefore necessary to coordinate the restoration of generators, network and loads to achieve a balance between the restorability and robustness.

### B. Optimization Modelling

It is hard for attackers to continuously monitor and interfere with the cyber system in the whole network reconfiguration process. Likewise, it is difficult for the dispatch center to accurately monitor the time and target of cyber-attacks. Thus, typical attacks against important and weak parts should be considered when formulating the preparatory schemes. The most possible attack targets [18] are transmission lines with the highest power flow [15] and their connected substations or power plants. Since attacking substations or power plants is equivalent to shutting down all their connected lines, attack targets are uniformly represented by lines in this study.

After cyber-attacks, the dispatch center should transit power grids to another stable operation state by reasonable control. It should be noted that subsequent restoration operations such as generator output rising, transmission line input and load pickup are unavailable in the transition process. In addition, studies in this paper are conducted based on the steady-state condition.

*1) Objective*

The resilience of the restoration state can be evaluated by the integral of the system performance function over the restoration period, which is defined as follows.

$$R = \int_{t_0}^{t_0+T} F(t)dt \quad (11)$$

where $R$ is the resilience metric in the period $[t_0, t_0+T]$, $T$ is the restoration time period, and $F(t)$ is the system performance function. As all the proposed resilience indices reflect a certain extent of load restoration, and the ultimate target of power system restoration is to restore all outage loads within a short time, the objective function is defined as:

$$\min R = \int_0^T [R_1(t) + R_2(t)]dt =$$

$$\int_0^T \left\{ [1-\eta(t)] \left[ 1 - \frac{\sum_{l \in L} P_{L,l}(t)}{P_{Load\_all}} \right] + \eta(t) \sum_{m=1}^M p(m) \frac{\sum_{l \in L} \Delta P_{L,l,m}(t)}{\sum_{l \in L} P_{L,l}(t)} \right\} dt \quad (12)$$

where $R$ is minimizing the sum of load loss after blackout and average secondary load curtailment caused by cyber-attacks. $R_1(t)$ is the restorability of schemes, which is equivalent to maximizing the restored load and minimizing the time consumption, and $R_2(t)$ is the robustness of restored power grids. At restoration time $t$, $\eta(t)$ is the percentage of the restored load to the total load. It indicates that with the increase of $\eta(t)$, the importance of the restorability gradually decreases and the robustness should be paid greater attention. The term $P_{Load\_all}$ is the total outage load, $p(m)$ is the occurrence probability of the $N$-$k$ contingency accident $m$ caused by cyber-attacks. Here $p(m)=\mu^k$, where $\mu$ is the fault probability of equipment, and $k$ is the number of equipment failures at the same time in incident $m$. $M$ is the number of anticipated incidents, and $\Delta P_{L,l,m}(t)$ is the load curtailment of load $l$ in incident $m$. Moreover, the decision variables are the restoration sequence of transmission lines.

In $R_1(t)$, the term $P_{L,l}(t)$ is obtained based on the optimal power flow (OPF) model [19]. In contrast with the standard OPF model, two additional constraints are added to calculate the $P_{L,l}(t)$ at each restored bus:

$$\begin{cases} P_{L,l}(t) \le P_{L,l}^0 \\ \sum_{l \in L} P_{L,l}(t) = \tau \sum_{g \in G} P_{G,g}(t) \end{cases} \quad (13)$$

where $P_{L,l}^0$ is the total load needed to be restored of load $l$, $\tau$ is a proportionality coefficient, $\tau \in [0,1]$, and $\sum_{g \in G} P_{G,g}(t)$ is the sum of the maximum output of generators. $P_{G,g}$ is calculated by the generator ramping function [9] as follows:

$$P_{G,g}(t) = \begin{cases} 0 & 0 \le t \le T_{s,g} \\ r_g(t - T_{s,g}) & T_{s,g} \le t \le T'_{s,g} \\ P_{GN,g} & t \ge T'_{s,g} \end{cases} \quad (14)$$

where $P_{GN,g}$ is the rated active power, $T_{s,g}$ and $T'_{s,g}$ are the start-up time and the time of reaching $P_{GN,g}$, $r_g$ is the ramping rate of generator $g$. In particular, the output of offline generators is 0.

In $R_2(t)$, $\Delta P_{L,l,m}(t) = \sum_{l \in L} [P_{L,l}(t) - P'_{L,l}(t)]$, where, $P'_{L,l}(t)$ is the remaining load after attacks, calculated by the following three-step process: 1) obtain all the initial and secondary fault equipment based on the attack propagation process; 2) search split sub-restored grids after cyber-attacks and; 3) calculate $P'_{L,l}(t)$ of each isolated sub-grids using the OPF method.

*2) Constraints*

*(1) Constraints of the restorability for the network reconfiguration model*

The start-up power constraint is that sufficient power should be supplied to support the NBSU start-up and load restoration. The sum of the cranking power of restarted generators and the active power of restored loads must be less than the power that the restored system can provide at time $t$, which is given by:

$$\sum_{g \in G} c_g(t) P_{cr,g}(t) + \sum_{l \in L} P_{L,l}(t) < \sum_{i \in G_{on}} P_{G,i}(t) \quad (15)$$

where $c_g(t)=1$ means generator $g$ has started, while 0 means not. $P_{G,i}(t)$ is the active power that generator $i$ can supply.

The time limitations are that thermal generators to be restarted have a maximum critical hot-start time limit $T_{CH,g}$ and a minimum critical cold-start time limit $T_{CC,g}$, which are:

$$\begin{cases} 0 < T_{s,g} < T_{CH,g} \\ T_{s,g} > T_{CC,g} \end{cases} \quad (16)$$

The system operation constraints mainly include the power balance constraint, the minimum and maximum system frequency, the upper or lower-power limits of generators, the bus voltage limit and the power flow limits of transmission lines. These constraints can be conceptually expressed by the following equations and inequalities.

$$\begin{cases} g(\boldsymbol{x}, \boldsymbol{u}) = 0 \\ q(\boldsymbol{x}, \boldsymbol{u}) < 0 \end{cases} \quad (17)$$

where *x* is the vector of power grid state variables, and *u* is the vector of power grid control variables.

*(2) Constraints of the robustness for the restored partial power grid with cyber-attack model*

After cyber-attacks, the redispatching resources can only be restored equipment, *e.g.* restored generators, loads and lines. Unrestored equipment is unavailable for security and stability control of power grids. Furthermore, there are some special constraints for calculating the OPF of isolated sub-grids. For maintaining the security and stability of the restored power grid, operators primarily aim to adjust the loads and generator output, and attempt to avoid artificially changing the network structure. First, the regulation range of the generator output is as follows.

$$0 \le P'_{G,g}(t) \le P_{G,g}(t) \qquad (18)$$

where $P'_{G,g}(t)$ is the output of generator $g$ after cyber-attacks and operator's regulation. The adjustable upper limit is $P_{G,g}(t)$. Second, the regulation range of the restored load is as follows.

$$0 \le P'_{L,l}(t) \le P_{L,l}(t) \qquad (19)$$

where $P'_{L,l}(t)$ is the remaining value of load $l$ after cyber-attacks and operator's regulation. The adjustable upper limit is $P_{L,l}(t)$.

### C. Solving Method

The network reconfiguration is a nonlinear mixed integer optimization problem. Furthermore, in the proposed model, only by a complex multi-step sequential simulation including the equipment shutdown, island searching and power flow calculation of split sub-restored grids, can $\Delta P_{L,l,m}(t)$ be obtained. It is hard to add this simulation process to the optimization problem by a mathematical description. Thus, the natural aggregation algorithm (NAA) [20] with a clear solution framework is used to solve the proposed model.

The NAA has very strong search capability and robustness compared with other evolutionary algorithms. In this study, each individual in NAA is first coded as an *L*-dimensional vector, where *L* is the total number of transmission lines to be restored. Second, it is decoded to the restored sequence of lines, which represents a network reconfiguration scheme. Third, the restorability model is calculated based on each scheme. Then, the propagation process of cyber-attacks is simulated and impacts are obtained. Finally, the robustness model with cyber-attacks is calculated. In summary, the solving process of resilient network reconfiguration is shown in Fig. 4.

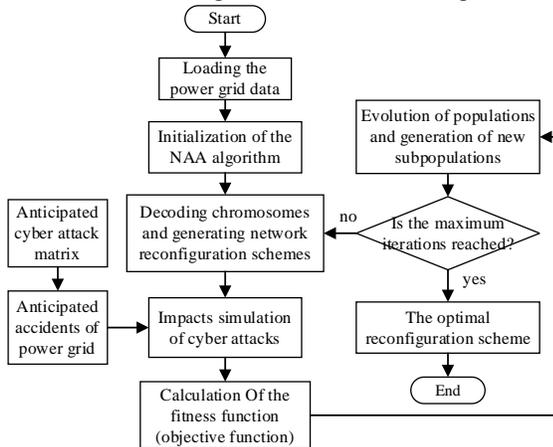

Fig. 4.  Flow chart of the optimization solving process of resilient network reconfiguration considering cyber-attacks.

## IV. CASE Study

### A. Set up

The IEEE 39-bus system is used to illustrate the proposed framework. It contains 10 generators, 46 transmission lines and the total load is 6254.23 MW. The generator connected to bus 30 is the BSU. The restoration-related parameters of generators are shown in Table I. The restoration time of a line is set as 5 minutes, and the time interval of a restoration step is also set as 5 minutes. Furthermore, charged outage loads or NBSUs will be put into operation in the next time step. When all NBSUs are restarted, the network reconfiguration is finished.

Generally, the attack ability and budgets of attackers are limited, and it is inefficient to carry out many cyber-attacks in one-time section. In this study, the number of failed equipment $k$ is set as 2 in *N-k* contingency. The probability of a single line fault $\mu$ is assumed to be 0.25. According to the power flow of transmission lines in descending order, the top 30% of lines are selected as the anticipated targets of cyber-attacks.

For the NAA algorithm, the population size and the generation are both set as 100. According to [20], the shelter number, the scaling factor, the local and global search crossover factor and the global movement amplification are set as 4, 1, 0.9, 0.1 and 1.2, respectively.

TABLE I
PARAMETERS OF GENERATORS IN THE IEEE 39-BUS SYSTEM

| Generator | $P_{GN}$ (MW) | $r_g$ (MW/h) |
|---|---|---|
| 30 | 1040.0 | 624.0 |
| 31 | 646.0 | 387.6 |
| 32 | 725.0 | 435.0 |
| 33 | 652.0 | 391.2 |
| 34 | 508.0 | 304.8 |
| 35 | 687.0 | 412.2 |
| 36 | 580.0 | 348.0 |
| 37 | 564.0 | 338.4 |
| 38 | 865.0 | 519.0 |
| 39 | 1100.0 | 660.0 |

### B. Optimal Network Reconfiguration Schemes

Two scenarios are considered, wherein N-1 faults and N-1 & N-2 faults caused by cyber-attacks are simulated, respectively. Table II shows the optimal schemes, recorded as **Scheme1** and **Scheme2**, corresponding to the two scenarios. Both schemes have the same network structure, shown in Fig.5(a), with the same restarted sequences of NBSUs, but the charging sequences of lines are different. The corresponding restoration time is 140min and 150min, respectively. It can be seen that generators closing to the BSU, with large rated power and fast ramping rate are restarted preferentially. Restoration paths of the two scenarios expand throughout the network rather than concentrated on a few lines. Moreover, the network structure of the two schemes both contain a loop network with lines ③, ④, ⑦, ㉚, ㉛, ㊵ and ㊷.

To illustrate the advantages of the proposed resilient network reconfiguration strategy, the typical traditional restoration strategy [21] using the shortest path algorithm is introduced for comparison. Reference [21]'s optimal network reconfiguration scheme is recorded as **Scheme3**. The network structure is shown in Fig.5(b), which is a tree graph. The comparison of the overall reconfiguration process of **Scheme1** and **Scheme2** and **Scheme3** is shown in Table II.

TABLE II
THE OPTIMAL NETWORK RECONFIGURATION SCHEMES

| Scheme1 | | | Scheme2 | | | Scheme3 | | |
|---|---|---|---|---|---|---|---|---|
| Generator restart sequence | Restoration time (min) | Charging sequence of transmission lines | Generator restart sequence | Restoration time (min) | Charging sequence of transmission lines | Generator restart sequence | Restoration time (min) | Charging sequence of transmission lines |
| 30 | 0 | -- | 30 | 0 | -- | 30 | 0 | -- |
| 39 | 15 | ⑤-①-② | 39 | 15 | ⑤-①-② | 32 | 35 | ⑤-③-⑥-⑨-㊳-⑲-⑳ |
| 37 | 25 | ④-㊶ | 37 | 25 | ④-㊶ | 35 | 65 | ⑦-㊵-㉒-㉓-㉖-㊲ |
| 38 | 40 | ㊵-㊹-㊻ | 38 | 40 | ㊵-㊹-㊻ | 31 | 80 | ⑧-⑩-⑭ |
| 33 | 65 | ㊷-㊱-㉖-㉗-㊳ | 33 | 65 | ㊷-㊱-㉖-㉗-㊳ | 37 | 90 | ④-㊶ |
| -- | 80 | ③-⑦-㊴ | -- | -- | -- | -- | -- | -- |
| 35 | 95 | ㉘-㉕-㉗ | 35 | 80 | ㉘-㉕-㉗ | 36 | 100 | ㊶-⑲ |
| -- | -- | -- | -- | 95 | ㊴-③-⑦ | -- | -- | -- |
| 36 | 105 | ㊶-⑲ | 36 | 105 | ㊶-⑲ | 34 | 115 | ㉗-㉜-㉔ |
| 34 | 115 | ㉜-㉔ | 34 | 115 | ㉜-㉔ | 38 | 130 | ㊵-㊹-㊻ |
| 31 | 135 | ⑥-⑧-⑩-⑭ | 31 | 135 | ⑥-⑧-⑩-⑭ | 39 | 140 | ①-② |
| 32 | 150 | ⑬-⑱-⑳ | 32 | 150 | ⑬-⑱-⑳ | 33 | 145 | ㊳ |

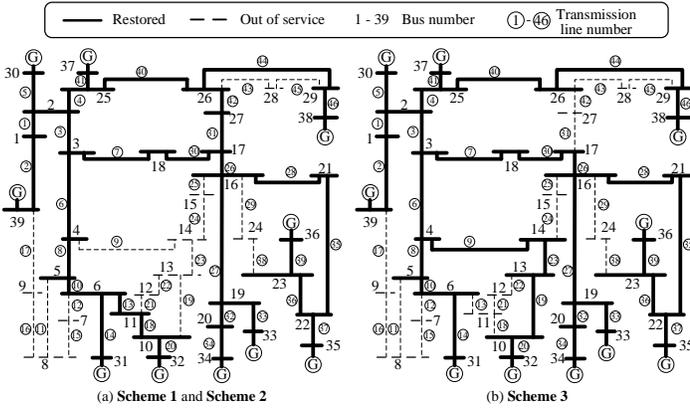

Fig. 5. The network structure of optimal network reconfiguration schemes.

(a) **Scheme 1** and **Scheme 2**     (b) **Scheme 3**

### C. Result Analysis

Table III presents a comparison of proposed restorability indices based on above three schemes. Table IV compares proposed robustness indices in several anticipated cyber-attack incidents: 1) The same cyber-attack leading to the outage of line ④ against each scheme occurs at 90 and 140 minutes, respectively; 2) Different cyber-attacks at the same time of 140 minutes leading to the shutdown of generator connected to bus 30 and the outage of bus 3, respectively.

TABLE III
RESTORABILITY INDEX OF OPTIMAL NETWORK RECONFIGURATION SCHEMES

| Reconfiguration scheme | Restorability indices | | | | |
|---|---|---|---|---|---|
| | $R_1^{res}$ (min) | $R_2^{res}$ (MW/h) | $R_3^{res}$ (MW) | $R_4^{res}$ | $R_5^{res}$ |
| **Scheme1** | 150 | 175.03 | 4648.80 | 0.7632 | 0.3077 |
| **Scheme2** | 150 | 180.19 | 4648.80 | 0.7632 | 0.3077 |
| **Scheme3** | 145 | 129.20 | 3642.85 | 0.7632 | 0.2935 |

As shown in Table III, except the restoration time is longer, the other four indices of **Scheme1** and **Scheme2** are all better than **Scheme3**. It shows that at the cost of increasing restoration time, restorability is improved by restoring several redundant transmission lines. An improved restoration effect can be obtained by cooperatively controlling the restoration of generators, power networks and loads.

At 140 min, the loop network has been restored both in **Scheme1** and **Scheme2**. The robustness indexes of the two restored grids are the same and only $R_5^{rob}$ slightly reduced, which are all better than **Scheme3**. Particularly, compared with the attacks at 90 minutes, $R_3^{rob}$, $R_4^{rob}$ and $R_5^{rob}$ of **Scheme2** are greatly improved, which means the impacts of cyber-attacks are greatly reduced. As for **Scheme3**, the impacts of outage of line ④ increase with the continuous restoration.

TABLE IV
ROBUSTNESS INDEX OF RESTORED GRIDS WITH SEVERAL CYBER-ATTACKS

| Incident caused by cyber-attack | Robustness indices | Scheme1 | Scheme2 | Scheme3 |
|---|---|---|---|---|
| Outage fault of transmission line ④ at 90 minutes | $R_1^{rob}$ (min) | 0 | 0 | 10 |
| | $R_2^{rob}$ (MW/h) | 0 | 0 | 338.4 |
| | $R_3^{rob}$ (%) | 0 | 14.94 | 12.33 |
| | $R_4^{rob}$ (%) | 0 | 44.44 | 28.89 |
| | $R_5^{rob}$ (%) | 12.07 | 40.94 | 20.47 |
| Outage fault of transmission line ④ at 140 minutes | $R_1^{rob}$ (min) | 0 | 0 | 10 |
| | $R_2^{rob}$ (MW/h) | 0 | 0 | 0 |
| | $R_3^{rob}$ (%) | 0 | 0 | 3.95 |
| | $R_4^{rob}$ (%) | 0 | 0 | 42.06 |
| | $R_5^{rob}$ (%) | 5.26 | 5.26 | 29.56 |
| Shutdown of generator 30 at 140 minutes | $R_1^{rob}$ (min) | 0 | 0 | 0 |
| | $R_2^{rob}$ (MW/h) | 624.0 | 624.0 | 624.0 |
| | $R_3^{rob}$ (%) | 22.15 | 21.04 | 30.63 |
| | $R_4^{rob}$ (%) | 11.11 | 11.11 | 12.5 |
| | $R_5^{rob}$ (%) | 0 | 0 | 0 |
| Outage fault of bus 3 at 140 minutes | $R_1^{rob}$ (min) | 15 | 15 | 20 |
| | $R_2^{rob}$ (MW/h) | 0 | 0 | 0 |
| | $R_3^{rob}$ (%) | 17.19 | 17.19 | 21.11 |
| | $R_4^{rob}$ (%) | 29.22 | 29.22 | 73.02 |
| | $R_5^{rob}$ (%) | 39.60 | 39.60 | 70.69 |

When the cyber-attacks result in the shutdown of generator 30 at 140 minutes, the robustness indices of these three schemes are similar, and the load curtailment of **Scheme3** is slightly larger. In particular, as the generator shutdown does not change the network structure, there is no change in $R_5^{rob}$.

In addition, the outage fault of bus 3 is a seriously anticipated incident. In terms of network structure, it is equivalent to the simultaneous outage of lines ③, ⑥ and ⑦. Fig. 7 shows the corresponding impact on network structure. After being attacked at 140 minutes, the restored power grids of the three schemes all split. Where the restored grids of **Scheme1** and **Scheme2** split into two sub-grids, and the shortest path consumes 15 minutes for reconnecting containing lines ⑨, ㉔ and ㉕, as shown in Fig. 7(a). The restored grid of **Scheme3** splits into three sub-grids, and the shortest path consuming 20 minutes for reconnecting contains lines ㉔, ㉕, ㉛ and ㊷, as shown in Fig. 7(b). $R_3^{rob}$, $R_4^{rob}$ and $R_5^{rob}$ of **Scheme1** and **Scheme2** are all better than **Scheme3**.

In conclusion, by using the proposed resilient network reconfiguration strategy, although the restoration time will increase, most resilience indices can be significantly improved. The resilient network reconfiguration scheme therefore effectively reduces secondary losses caused by cyber-attacks.

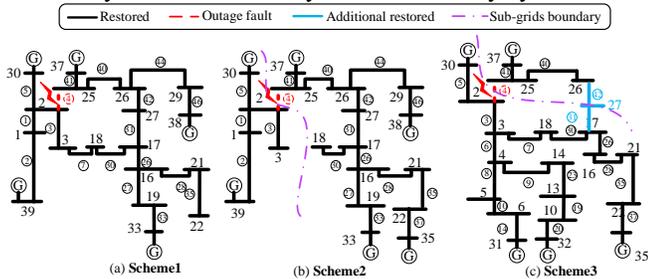

Fig. 6. Outage impact on network structure of line ④ at 90 minutes.

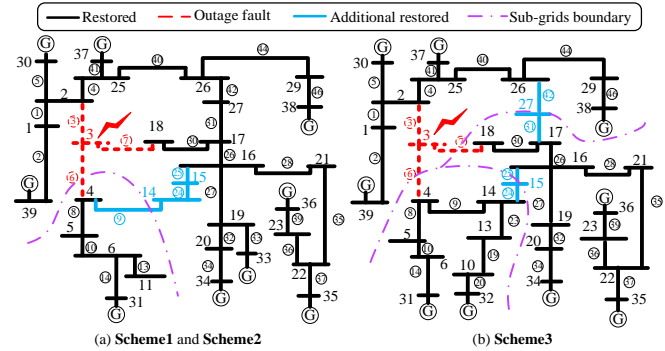

Fig. 7. Outage impact on network structure of bus 3 at 140 minutes.

## V. CONCLUSION

Considering the impacts and propagation process of cyber-attacks, this paper proposes a framework of resilient transmission network reconfiguration. From the perspective of the control center, the proposed strategy improves power grid resilience in both restorability and robustness by optimally controlling the restoration process on the physical side. Based on typical incidents of *N-k* contingency caused by anticipated cyber-attacks, an optimization model of resilient network reconfiguration is proposed. Simulation results show that compared with the traditional strategy, the proposed framework can effectively improve the power grid resilience and defend against cyber-attacks in the network reconfiguration process.

Further detailed research will focus on 1) the rationale behind the cyber-attack vector establishment including the attack time and position considering the characteristics of power system restoration. 2) How to launch more reasonable attacks to simulate realistic impacts. 3) Extending the scope to the entire restoration process of the CPPS.